\documentclass[12pt,twocolumn,english]{IEEEtran}
\usepackage[T1]{fontenc}
\usepackage[latin9]{inputenc}
\usepackage{geometry}
\usepackage{float}
\usepackage{amsmath}
\usepackage{graphicx}
\usepackage{setspace}
\usepackage{amssymb}

\makeatletter

\providecommand{\tabularnewline}{\\}

\onecolumn{}
\date{}
\usepackage{cite}

\makeatother

\usepackage{babel}

\begin{document}

\title{Frequency Domain Hybrid--ARQ Chase Combining for Broadband MIMO CDMA
Systems%
\thanks{\noindent This work was partly supported by Maroc Telecom under contract
105 10005462.06/PI. This paper was presented in parts at IEEE IWCMC2008,
Crete Island, Greece, Aug. 2008, and IEEE PIMRC 2008, Cannes, France,
September 2008.\protect \\
H. Chafnaji (houda.chafnaji@telecom-bretagne.eu) and T. Ait-Idir (aitidir@ieee.org)
are with the Communications Systems Department, INPT, Madinat Al-Irfane,
Rabat, Morocco. They are also with the Signal and Communications Department,
Institut Telecom/Telecom Bretagne, Brest, France. S. Saoudi (samir.saoudi@telecom-bretagne.eu)
is with the Signal and Communications Department, Institut Telecom/Telecom
Bretagne, Brest, France. Athanasios V. Vasilakos (vasilako@ath.forthnet.gr)
is with the University of Western Macedonia, Greece. %
}}

\author{Houda Chafnaji{\normalsize ,} Tarik Ait-Idir{\normalsize ,} \emph{Member,
IEEE,}{\normalsize{} }and Samir Saoudi\textit{,} \emph{Member, IEEE,
}and Athanasios V. Vasilakos}

\maketitle
\begin{abstract}
In this paper, we consider high-speed wireless packet access using
code division multiple access (CDMA) and multiple-input--multiple-output
(MIMO). Current wireless standards, such as high speed packet access
(HSPA), have adopted multi-code transmission and hybrid--automatic
repeat request (ARQ) as major technologies for delivering high data
rates. The key technique in hybrid--ARQ, is that erroneous data packets
are kept in the receiver to detect/decode retransmitted ones. This
strategy is refereed to as \emph{packet combining}. In CDMA MIMO-based
wireless packet access, multi-code transmission suffers from severe
performance degradation due to the loss of code orthogonality caused
by both interchip interference (ICI) and co-antenna interference (CAI).
This limitation results in large transmission delays when an ARQ mechanism
is used in the link layer. In this paper, we investigate efficient
minimum mean square error (MMSE) frequency domain equalization (FDE)-based
iterative (turbo) packet combining for cyclic prefix (CP)-CDMA MIMO
with Chase-type ARQ. We introduce two turbo packet combining schemes:
i) In the first scheme, namely \emph{{}``chip-level turbo packet
combining''}, MMSE FDE and packet combining are jointly performed
at the chip-level. ii) In the second scheme, namely \emph{{}``symbol-level
turbo packet combining''}, chip-level MMSE FDE and despreading are
separately carried out for each transmission, then packet combining
is performed at the level of the soft demapper. The computational
complexity and memory requirements of both techniques are quite insensitive
to the ARQ delay, i.e., maximum number of ARQ rounds. The throughput
is evaluated for some representative antenna configurations and load
factors (i.e., number of orthogonal codes with respect to the spreading
factor) to show the gains offered by the proposed techniques. 
\end{abstract}

\section{Introduction}

Space-time (ST) multiplexing oriented multiple-input--multiple-output
(MIMO) and hybrid--automatic repeat request (ARQ) are two core technologies
used in the emerging code division multiple access (CDMA)-based wireless
packet access standards \cite{EricssonVTCS07}. In ST multiplexing
architectures, independent data streams are sent over multiple antennas
to increase the transmission rate \cite{Wolniansky}. In hybrid--ARQ,
erroneous data packets are kept in the receiver to help decode the
retransmitted packet, using \emph{packet combining} techniques (e.g.
see \cite{Harbey_Vicker_TCOM94} and references therein). 

To support heterogeneous data rates in CDMA systems, multiple spreading
codes can simultaneously be allocated to the same user if he requests
a high data rate \cite{CL-RDG}. This method is often refereed to
as \emph{{}``multi-code transmission,''} and has been considered
in the high speed packet access (HSPA) system \cite{HSPA_SpecR7}.
In MIMO CDMA systems, multi-code transmission offers a spectrum efficiency
that linearly increases in the order of the number of spreading codes
and transmit antennas. This is achieved by assigning the same spreading
code group to all transmit antennas. However, in severe frequency
selective fading wireless channels, the performance of this scheme
can dramatically deteriorate due to co-antenna interference (CAI)
and inter-chip interference (ICI). This results in a large delay (due
to multiple transmissions) when an ARQ protocol is used in the link
layer. Motivated by this limitation, we investigate efficient hybrid--ARQ
receiver schemes that allow to reduce the number of ARQ rounds required
to correctly decode a data packet in MIMO CDMA ARQ systems with multi-code
transmission.

Recently, cyclic-prefix (CP) aided single carrier (SC) CDMA transmission
with chip-level minimum mean square error (MMSE)-based frequency domain
equalization (FDE) has been introduced \cite{Adachi}. It is a transceiver
scheme that allows to achieve attractive performance with affordable
computational complexity cost. Turbo MMSE-FDE for CP-CDMA has then
been proposed to cope with severe ICI \cite{Lee}. In \cite{Garg_Adachi_JSAC_06},
MMSE FDE has been applied to perform packet combining for multi-code
CP-CDMA systems with ARQ operating over severe frequency selective
fading channels. It has recently been demonstrated that ARQ presents
an important source of diversity in MIMO systems \cite{Elgamal}.
Interestingly, it has been shown in \cite{Elgamal} that for both
short and long-term static %
\footnote{The short-term static ARQ channel dynamic corresponds to the case
where two consecutive ARQ rounds observe independent channel realizations.
In long-term static channels, all ARQ rounds corresponding to the
same data packet observe the same channel realization. %
} ARQ channel dynamics, multiple transmissions improve the diversity
order of the corresponding MIMO ARQ channel. The case of block-fading
MIMO ARQ, i.e., multiple fading blocks are observed within the same
ARQ round, has been reported in \cite{Collings_DMD_BlockFadingIT08}.
Information rates and turbo MMSE packet combining strategies for frequency
selective fading MIMO ARQ channel have been investigated in \cite{Ait-Idir_SaoudiTCOM08}.
Turbo MMSE packet combining for broadband MIMO ARQ systems with co-channel
interference (CCI) has recently been reported in \cite{Ait-Idir_PIMRC08}
and \cite{Ait-Idir_TWC_Mar09} using time and frequency domain combining
methods, respectively. 

In this paper, we consider Chase-type ARQ with multi-code CP-CDMA
MIMO transmission %
\footnote{In this MIMO CDMA ARQ transmission scheme, the chip packet is completely
retransmitted at each ARQ round.%
} over broadband wireless channel. We propose two iterative (turbo)
packet combining schemes where, at each ARQ round, the data packet
is decoded by iteratively exchanging soft information in the form
of log-likelihood ratios (LLRs) between the soft-input--soft-output
(SISO) packet combiner and the SISO decoder. In the first turbo packet
combining scheme, we exploit the fact that both the CP chip-word and
data packet are retransmitted at each ARQ round. This allows us to
view each transmission as a group of virtual receive antennas, and
build up a virtual MIMO channel that takes into account both multi-antenna
and multi-round transmission. We therefore perform combining of multiple
transmissions jointly with chip-level soft MMSE FDE. This scheme is
called \emph{chip-level packet combining}. In the second scheme, both
chip-level soft MMSE FDE and despreading are separately carried out
for each transmission. Combining is then performed at the level of
the soft symbol demapper. We analyze both the computational complexity
and memory required by the proposed techniques, and show that they
are less sensitive to the ARQ delay, i.e., maximum number of ARQ rounds.
Finally, we evaluate and compare the throughput performance of the
proposed schemes for some representative load factors (i.e., number
of parallel codes with respect to the spreading factor) and antenna
configurations. 

Throughout this paper, $\left(.\right)^{\top}$ and $\left(.\right)^{\mathrm{H}}$
denote the transpose and transpose conjugate of the argument, respectively.
$\mathrm{diag}\left\{ \mathbf{x}\right\} \in\mathbb{C}^{n\times n}$
and $\mathrm{diag}\left\{ \mathbf{X}_{1},\cdots,\mathbf{X}_{m}\right\} \in\mathbb{C}^{mn_{1}\times mn_{2}}$
denote the diagonal matrix and block diagonal matrix constructed from
$\mathbf{x}\in\mathbb{C}^{n}$ and $\mathbf{X}_{1},\cdots,\mathbf{X}_{m}\in\mathbb{C}^{n_{1}\times n_{2}}$,
respectively. For $\mathbf{x}\in\mathbb{C}^{TN}$, $\mathbf{x}_{f}$
denotes the discrete Fourier transform (DFT) of $\mathbf{x}$, i.e.
$\mathbf{x}_{f}=\mathbf{U}_{T,N}\mathbf{x}$, with $\mathbf{U}_{T,N}=\mathbf{U}_{T}\otimes\mathbf{I}_{N}$,
where $\mathbf{I}_{N}$ is the $N\times N$ identity matrix, $\mathbf{U}_{T}$
is a unitary $T\times T$ matrix whose $\left(m,n\right)$th element
is $\left(\mathbf{U}_{T}\right)_{m,n}=\frac{1}{\sqrt{T}}e^{-j(2\pi mn/T)}$,
$j=\sqrt{-1}$, and $\otimes$ denotes the Kronecker product. The
rest of this paper has the following structure. In Section \ref{sec:SystemDescription},
we present the CP-CDMA MIMO ARQ transmission scheme then provide its
corresponding communication model. In Section \ref{sec:Iterative-Receivers},
we derive the two iterative soft MMSE FDE-aided packet combining schemes
we propose in this paper. Section \ref{sec:Complexity-and-Performance},
analyzes the complexity and memory size required by both schemes,
then focuses on the comparison of their throughput performances. The
paper is concluded in Section \ref{sec:Conclusions}. \vspace{-4mm}

\section{System Description\label{sec:SystemDescription}\vspace{-3mm}
}

\subsection{CP-CDMA MIMO ARQ Transmission Scheme}

We consider a single user multi-code CP-CDMA transmission scheme over
a broadband MIMO channel with an ARQ protocol in the upper layer,
where the ARQ delay is $K$ (index $k=1,\cdots,K$). An information
block is first encoded using a $\rho$-rate encoder, then interleaved
with the aid of a semi-random interleaver $\Pi$, and spatially multiplexed
over $N_{T}$ transmit antennas (index $t=1,\cdots,N_{T}$) to produce
the coded and interleaved frame $\boldsymbol{b}$ which is \emph{serial-to-parallel}
converted to $N_{T}$ sub-streams $\boldsymbol{b}_{1},\ldots,\boldsymbol{b}_{N_{T}}$
, where\begin{equation}
\mathbf{b}_{t}\triangleq\left[b_{t,0,1},\cdots,b_{t,j,m},\cdots,b_{t,T_{s}-1,M}\right]\in\left\{ 0,1\right\} ^{MT_{S}}.\label{eq:coded and inteleaved bits}\end{equation}

\noindent $T_{s}$ denotes the length of the symbol block transmitted
over each antenna (index $j=0,\cdots,T_{s}-1$). Each sub-stream is
then symbol mapped onto the elements of constellation $\mathcal{S}$
where $\left|\mathcal{S}\right|=2^{M}$. For each antenna, the symbol
block is passed through a \emph{serial-to-parallel} converter and
a spreading module which consists in $C$ orthogonal codes. The same
spreading matrix\vspace{-2mm}
 \begin{equation}
\mathbf{W}\triangleq\left[\mathbf{w}_{1}^{\top},\cdots,\mathbf{w}_{C}^{\top}\right]\in\left\{ \pm1/\sqrt{N}\right\} ^{N\times C}\label{eq:Spreadingmatrix}\end{equation}
 is used for each transmit antenna, where\vspace{-2mm}
 \begin{equation}
\mathbf{w}_{n}\triangleq\left[w_{1,n},\cdots,w_{N,n}\right],\,\, n=1,\cdots,C,\label{eq:Wlash_code}\end{equation}
 is a Walsh code of length $N$ (i.e., spreading factor), and $C\leq N$
is the number of multiplexed codes. The rate of this space-time code
(STC) is therefore\vspace{-2mm}
 \begin{equation}
R=\rho MN_{T}C.\label{eq:Rate_Transmitter}\end{equation}
The $C$ parallel chip-streams on each antenna are then added together
to construct a block of $T_{c}=T_{s}\frac{N}{C}$ chips (index $i=0,\cdots,T_{c}-1$).
The chips at the output of the $N_{T}$ transmit antennas are arranged
in the $N_{T}\times T_{c}$ matrix \begin{align}
\mathbf{X} & \triangleq\begin{array}{ccccc}
\biggl\lceil & x_{1,0} & \cdots & x_{1,T_{c}-1} & \biggr\rceil\\
\biggl| & \vdots &  & \vdots & \biggr|\\
\biggl\lfloor & \underbrace{x_{N_{T},0}} & \cdots & \underbrace{x_{N_{T},T_{c}-1}} & \biggr\rfloor\\
 & \mathbf{x}_{0} &  & \mathbf{x}_{T_{c}-1}\end{array},\label{eq:CHip_Matrix}\end{align}
where \begin{equation}
x_{t,i}\triangleq{\displaystyle \sum_{n=1}^{C}s_{t,n,i}w_{p,n},\,\,\, p=i\,\mathrm{mod}\, N+1},\label{eq:antenna_output}\end{equation}
and $s_{t,n,i}$ denotes the symbol transmitted by antenna $t$ at
channel use (c.u) $i$ using Walsh code $\mathbf{w}_{n}$. Transmitted
chips are independent (infinitely deep interleaving assumption), and
the chip energy is normalized to one, i.e., $\mathbb{{E}}\left[\left|x_{t,i}\right|^{2}\right]=1$
. A CP chip-word of length $T_{CP}$ is appended to $\mathbf{X}$
to construct the $N_{T}\times\left(T_{c}+T_{CP}\right)$ chip matrix
$\mathbf{X}'$ to be transmitted. We consider Chase-type ARQ: When
the decoding outcome is erroneous at ARQ round $k$, the receiver
feeds back a negative acknowledgment (NACK) message, then the transmitter
completely retransmits chip-matrix $\mathbf{X}'$ in the next round.
A successful decoding incurs the feed back of a positive acknowledgment
(ACK) message. The transmitter then stops the transmission of the
current frame and moves on to the next frame. Fig. \ref{fig:STBICM_ARQ}
depicts the considered CP-CDMA MIMO transmission scheme with ACK/NACK.

\subsection{Communication Model}

The broadband MIMO propagation channel connecting the $N_{T}$ transmit
and the $N_{R}$ receive antennas is composed of $L$ chip-spaced
taps (index $l=0,\cdots,L-1$). We assume a quasi-static block fading
channel, i.e., the channel is constant over an information block and
independently changes from block to block. The $N_{R}\times N_{T}$
channel matrix characterizing the $l$th discrete tap at ARQ round
$k$ is denoted $\mathbf{H}_{l}^{\left(k\right)}$, and is made of
zero-mean circularly symmetric complex Gaussian random entries. The
average channel energy per receive antenna is normalized as \begin{equation}
{\displaystyle \sum_{l=0}^{L-1}}{\displaystyle \sum_{t=1}^{N_{T}}}\,\,\mathbb{{E}}\left[\left|h_{r,t,l}^{\left(k\right)}\right|^{2}\right]=N_{T},\,\, r=1,\cdots,N_{R},\label{eq:Ch_energy_Normal}\end{equation}
 where $h_{r,t,l}^{\left(k\right)}$ is the $\left(r,t\right)$th
element of $\mathbf{H}_{l}^{\left(k\right)}$. 

At the receiver side, after removing the CP-word at ARQ round $k$,
a DFT is applied on received signals. This yields $T_{c}$ frequency
domain components grouped in block \begin{equation}
\mathbf{y}_{f}^{\left(k\right)}\triangleq\left[\mathbf{y}_{f_{0}}^{\left(k\right)^{\top}},\cdots,\mathbf{y}_{f_{T_{c}-1}}^{\left(k\right)^{\top}}\right]^{\top},\label{eq:FD_Rx_Signal}\end{equation}
which can be expressed as, \begin{equation}
\mathbf{y}_{f}^{\left(k\right)}=\boldsymbol{{\Lambda}}^{\left(k\right)}\mathbf{x}_{f}+\begin{alignedat}{1}\mathbf{n}_{f}^{\left(k\right)}\end{alignedat}
,\label{eq:FD_CommModel1}\end{equation}
where vectors \begin{equation}
\mathbf{x}_{f}\triangleq\left[\mathbf{x}_{f_{0}}^{\top},\cdots,\mathbf{x}_{f_{T_{c}-1}}^{\top}\right]^{\top}\in\mathbb{C}^{T_{c}N_{T}\times1},\label{eq:FD_ChipVector}\end{equation}
\begin{equation}
\mathbf{n}_{f}^{\left(k\right)}\triangleq\left[\mathbf{n}_{f_{0}}^{\left(k\right)^{\top}},\cdots,\mathbf{n}_{f_{T_{c}-1}}^{\left(k\right)^{\top}}\right]^{\top},\label{eq:FD_Noise}\end{equation}
group the DFTs of transmitted chips and thermal noise at round $k$,
respectively, and $\mathbf{n}_{f}^{\left(k\right)}\sim\mathcal{N}\left(\mathbf{0},\sigma^{2}\mathbf{I}_{T_{c}N_{R}}\right)$.
The channel frequency response (CFR) matrix $\boldsymbol{{\Lambda}}^{\left(k\right)}$
at ARQ round $k$ is given by

\begin{equation}
\left\{ \begin{array}{l}
\boldsymbol{{\Lambda}}^{\left(k\right)}\triangleq\mathrm{diag}\left\{ \boldsymbol{{\Lambda}}_{0}^{\left(k\right)},\cdots,\boldsymbol{{\Lambda}}_{T_{c}-1}^{\left(k\right)}\right\} ,\\
\boldsymbol{{\Lambda}}_{i}^{\left(k\right)}=\sum_{l=0}^{L-1}\mathbf{H}_{l}^{\left(k\right)}e^{-j(2\pi il/T_{c})}.\end{array}\right.\label{eq:Lambda_Eqs}\end{equation}

\section{Iterative Receivers for CP-CDMA MIMO ARQ\label{sec:Iterative-Receivers}}

In this section, we present two efficient algorithms for performing
turbo packet combining for CP-CDMA MIMO ARQ systems : i) chip-level
turbo packet combining, and ii) symbol-level turbo packet combining.
In both schemes, signals received in multiple ARQ rounds are processed
using soft MMSE FDE. Transmitted data blocks are decoded, at each
ARQ round, in an iterative fashion through the exchange of soft information,
in the form of LLR values, between the soft packet combiner, i.e.,
soft--over ARQ rounds equalizer and demapper, and SISO decoder.

\subsection{Chip-Level Turbo Packet Combining}

To exploit the diversity available in received signals $\mathbf{y}_{f_{0}}^{\left(1\right)},\cdots,\mathbf{y}_{f_{T_{c}-1}}^{\left(k\right)}$,
we view each ARQ round $k$ as an additional group of virtual $N_{R}$
receive antennas. The MIMO ARQ system can therefore be considered
as a point-to-point MIMO link with $N_{T}$ transmit and $kN_{R}$
receive antennas, where the $T_{c}kN_{R}\times1$ chip-level virtual
received signal vector $\underline{\mathbf{y}}_{f}^{(k)}$ is constructed
as,

\begin{equation}
\underline{\mathbf{y}}_{f}^{\left(k\right)}\triangleq\left[\mathbf{y}_{f_{0}}^{\left(1\right)^{\mathrm{\top}}},\cdots,\mathbf{y}_{f_{0}}^{\left(k\right)^{\mathrm{\top}}},\cdots,\mathbf{y}_{f_{T_{c}-1}}^{\left(1\right)^{\mathrm{\top}}},\cdots,\mathbf{y}_{f_{T_{c}-1}}^{\left(k\right)^{\mathrm{\top}}}\right]^{\mathrm{\top}}.\label{eq:HARQ_receiverIn}\end{equation}
 The frequency domain communication model after $k$ rounds is then
given as,

\begin{equation}
\underline{\mathbf{y}}_{f}^{\left(k\right)}=\underline{\boldsymbol{{\Lambda}}}^{\left(k\right)}\mathbf{x}_{f}+\begin{alignedat}{1}\underline{\mathbf{n}}_{f}^{\left(k\right)}\end{alignedat}
,\label{eq:ARQ_FD_CommModel}\end{equation}
where \begin{equation}
\underline{\boldsymbol{{\Lambda}}}^{\left(k\right)}\triangleq\mathrm{diag}\left\{ \left[\begin{array}{c}
\boldsymbol{{\Lambda}}_{0}^{\left(1\right)}\\
\vdots\\
\boldsymbol{{\Lambda}}_{0}^{\left(k\right)}\end{array}\right],\cdots,\left[\begin{array}{c}
\boldsymbol{{\Lambda}}_{T_{c}-1}^{\left(1\right)}\\
\vdots\\
\boldsymbol{{\Lambda}}_{T_{c}-1}^{\left(k\right)}\end{array}\right]\right\} \in\mathbb{C}^{T_{c}kN_{R}\times T_{c}N_{T}},\label{eq:lambda}\end{equation}
 and \begin{equation}
\begin{alignedat}{1}\underline{\mathbf{n}}_{f}^{\left(k\right)}\end{alignedat}
=\left[\mathbf{n}_{f_{0}}^{\left(1\right)^{\mathrm{\top}}},\cdots,\mathbf{n}_{f_{0}}^{\left(k\right)^{\mathrm{\top}}},\cdots,\mathbf{n}_{f_{T_{c}-1}}^{\left(1\right)^{\mathrm{\top}}},\cdots,\mathbf{n}_{f_{T_{c}-1}}^{\left(k\right)^{\mathrm{\top}}}\right]^{\mathrm{\top}}.\label{eq:noise}\end{equation}

Soft ICI cancellation and frequency domain MMSE filtering are jointly
performed over all ARQ rounds. We call this concept \emph{{}``chip-level
turbo packet combining}''. This requires a huge computational cost
since the complexity of computing MMSE filters is cubic in the order
of the ARQ delay. In addition, the required receiver memory size linearly
scales with the ARQ delay because all CFRs $\boldsymbol{{\Lambda}}_{0}^{\left(1\right)},\cdots,\boldsymbol{{\Lambda}}_{T_{c}-1}^{\left(k\right)}$
are required at round $k$ \cite{chafnaji_PIMRC08}. In the following,
we introduce an efficient turbo MMSE implementation algorithm for
chip-level combining where both receiver complexity and memory requirements
are quite insensitive to the ARQ delay. 

Let $\tilde{\mathbf{x}}$ and $\sigma_{t,i}^{2}$ denote the conditional
mean and variance of $\mathbf{x}$ and $x_{t,i}$, respectively. Soft
MMSE processing can be written in a compact forward-backward filtering
structure as in \cite{Visoz_TCOM06}. By using the matrix inversion
lemma \cite{Haykin_AdaptiveFiltering}, we can express soft MMSE chip-level
packet combining at round $k$ as, 

\begin{equation}
\mathbf{z}_{f}^{\left(k\right)}=\boldsymbol{{\Gamma}}^{\left(k\right)}\underline{\mathbf{\tilde{y}}}_{f}^{(k)}-\boldsymbol{{\Omega}}^{\left(k\right)}\tilde{\mathbf{x}}_{f},\label{eq:MMSE_plus_IC_metric_FD}\end{equation}
where $\boldsymbol{{\Gamma}}^{\left(k\right)}=\mathrm{diag}\left\{ \boldsymbol{{\Gamma}}_{0}^{\left(k\right)},\cdots,\boldsymbol{{\Gamma}}_{T_{c}-1}^{\left(k\right)}\right\} \in\mathbb{C}^{T_{c}N_{T}\times T_{c}N_{T}}$,
and $\boldsymbol{{\Omega}}^{\left(k\right)}=\mathrm{diag}\left\{ \boldsymbol{{\Omega}}_{0}^{\left(k\right)},\cdots,\boldsymbol{{\Omega}}_{T_{c}-1}^{\left(k\right)}\right\} \in\mathbb{C}^{T_{c}N_{T}\times T_{c}N_{T}}$
denote the forward and backward filters at round $k$, respectively,
and are given by,

\begin{equation}
\left\{ \begin{array}{l}
\boldsymbol{{\Gamma}}_{i}^{\left(k\right)}\triangleq\frac{1}{\sigma^{2}}\left\{ \mathrm{\mathbf{I}}_{N_{T}}-\underline{\mathbf{D}}_{i}^{(k)}\mathbf{C}_{i}^{\left(k\right)^{-1}}\right\} ,\\
\mathbf{C}_{i}^{\left(k\right)}=\sigma^{2}\boldsymbol{\tilde{{\Xi}}}^{-1}+\underline{\mathbf{D}}_{i}^{(k)},\end{array}\right.\label{eq:forward_filter}\end{equation}
\begin{equation}
\left\{ \begin{array}{l}
\boldsymbol{{\Omega}}_{i}^{\left(k\right)}\triangleq\boldsymbol{{\Gamma}}_{i}^{\left(k\right)}\underline{\mathbf{D}}_{i}^{(k)}-\boldsymbol{{\Upsilon}}^{\left(k\right)},\\
\boldsymbol{{\Upsilon}}^{\left(k\right)}=\frac{1}{T}{\displaystyle \sum_{i=0}^{T-1}}\,\boldsymbol{{\Gamma}}_{i}^{\left(k\right)}\underline{\mathbf{D}}_{i}^{(k)}.\end{array}\right.\label{eq:backward_filter}\end{equation}
$\boldsymbol{\tilde{{\Xi}}}$ is the $N_{T}\times N_{T}$ unconditional
covariance of transmitted chips, and is computed as the time average
of conditional covariance matrices $\boldsymbol{{\Xi}}_{i}\triangleq\mathrm{diag}\left\{ \sigma_{1,i}^{2},\cdots,\sigma_{N_{T},i}^{2}\right\} $.
Variables $\underline{\mathbf{\tilde{y}}}_{f}^{(k)}$ and $\underline{\mathbf{D}}_{i}^{(k)}$
are computed according to the following recursions, \begin{equation}
\left\{ \begin{array}{l}
\underline{\mathbf{\tilde{y}}}_{f}^{(k)}=\underline{\mathbf{\tilde{y}}}_{f}^{(k-1)}+\boldsymbol{{\Lambda}}^{\left(k\right)^{H}}\mathbf{y}_{f}^{\left(k\right)},\\
\underline{\mathbf{\tilde{y}}}_{f}^{(0)}=\boldsymbol{0}_{T_{c}N_{T}\times1},\end{array}\right.\label{eq:y_update}\end{equation}
\begin{equation}
\left\{ \begin{array}{l}
\underline{\mathbf{D}}_{i}^{(k)}=\underline{\mathbf{D}}_{i}^{(k-1)}+\boldsymbol{{\Lambda}}_{i}^{\left(k\right)^{H}}\boldsymbol{{\Lambda}}_{i}^{\left(k\right)},\\
\underline{\mathbf{D}}_{i}^{(0)}=\boldsymbol{0}_{N_{T}\times N_{T}}.\end{array}\right.\label{eq:D_update}\end{equation}
Note that recursions (\ref{eq:y_update}) and (\ref{eq:D_update})
present an important ingredient in the proposed chip-level combining
algorithm since both complexity and memory requirements become less
sensitive to the ARQ delay. These issues are discussed in detail in
Section \ref{sec:Complexity-and-Performance}. The inverse DFT (IDFT)
is then applied to $\mathbf{z}_{f}^{\left(k\right)}$ to obtain the
equalized time domain chip sequence. After despreading, extrinsic
LLR value $\phi_{t,j,m}^{(e)}$ corresponding to coded and interleaved
bit $b_{t,j,m}$ $\forall\,\, t,j,m$ is computed as,

\begin{equation}
\phi_{t,j,m}^{(e)}=\log\frac{{\displaystyle \sum_{s\in\mathcal{S}_{1}^{m}}\exp\left\{ \boldsymbol{\xi}_{t,j}^{(k)}(s)+{\displaystyle \sum_{m'\neq m}}\phi_{t,j,m'}^{(a)}\lambda_{m'}\left\{ s\right\} \right\} }}{{\displaystyle \sum_{s\in\mathcal{S}_{0}^{m}}\exp\left\{ \boldsymbol{\xi}_{t,j}^{(k)}(s)+{\displaystyle \sum_{m'\neq m}}\phi_{t,j,m'}^{(a)}\lambda_{m'}\left\{ s\right\} \right\} }},\label{eq:Ext_LLR}\end{equation}
where {\small $\boldsymbol{\xi}_{t,j}^{(k)}(s)=\frac{\left|r_{t,j}^{(k)}-g_{t,j}^{(k)}s\right|^{2}}{\theta_{t,j}^{(k)^{2}}}$},
with $r_{t,j}^{(k)}$, $g_{t,j}^{(k)}$, and $\theta_{t,j}^{(k)^{2}}$
are the despreading module output, gain, and residual interference
variance, respectively. $\phi_{t,j,m'}^{(a)}$ denotes \textit{a-priori}
LLR value corresponding to $b_{t,j,m'}$. $\lambda_{m'}\left\{ s\right\} $
is an operator that allows to extract the $m'$th bit labeling symbol
$s\in\mathcal{S}$, and $\mathcal{S}_{\beta}^{m}$ is the set of symbols
where the $m$th bit is equal to $\beta$, i.e. $\mathcal{S}_{\beta}^{m}=\left\{ s\,:\,\lambda_{m}\left\{ s\right\} =\beta\right\} $.
The obtained extrinsic LLR values are de-interleaved and fed to the
SISO decoder. The proposed low complexity algorithm is summarized
in Table \ref{tab:ITEQ_algorithm}.

\subsection{Symbol-Level Turbo Packet Combining}

In this combining scheme, the receiver performs chip-level space-time
frequency domain equalization separately for each ARQ round, then
combines multiple transmissions at the level of the soft demapper.
At each iteration of ARQ round $k$, soft ICI cancellation and MMSE
filtering are performed similarly to (\ref{eq:MMSE_plus_IC_metric_FD})
using communication model (\ref{eq:FD_CommModel1}). Extrinsic information
is computed using despreading module outputs corresponding to all
ARQ rounds. This requires the inversion of the $k\times k$ covariance
matrix of residual interference plus noise. By observing that despreading
module outputs obtained at different transmissions are independent,
extrinsic LLR value $\phi_{t,j,m}^{(e)}$ corresponding to coded and
interleaved bit $b_{t,j,m}$ can be expressed as, \begin{equation}
\phi_{t,j,m}^{(e)}=\log\frac{{\displaystyle \sum_{s\in\mathcal{S}_{1}^{m}}\exp\left\{ \boldsymbol{\overline{\xi}}_{t,j}^{(k)}(s)+{\displaystyle \sum_{m'\neq m}}\phi_{t,j,m'}^{(a)}\lambda_{m'}\left\{ s\right\} \right\} }}{{\displaystyle \sum_{s\in\mathcal{S}_{0}^{m}}\exp\left\{ \boldsymbol{\overline{\xi}}_{t,j}^{(k)}(s)+{\displaystyle \sum_{m'\neq m}}\phi_{t,j,m'}^{(a)}\lambda_{m'}\left\{ s\right\} \right\} }},\label{eq:Symbol_Comb}\end{equation}
where $\boldsymbol{\overline{\xi}}_{t,j}^{(k)}(s)$ is recursively
computed according to the following recursion, \begin{equation}
\left\{ \begin{array}{l}
\boldsymbol{\overline{\xi}}_{t,j}^{(k)}(s)=\boldsymbol{\overline{\xi}}_{t,j}^{(k-1)}(s)+\frac{\left|r_{t,j}^{(k)}-g_{t,j}^{(k)}s\right|^{2}}{\theta_{t,j}^{(k)^{2}}},\\
\boldsymbol{\overline{\xi}}_{t,j}^{(0)}(s)=0.\end{array}\right.\label{eq:LLR_Comb}\end{equation}
Note that this recursive implementation relaxes both the complexity
and memory requirements. The proposed low complexity algorithm is
summarized in Table \ref{tab:Symbol_level_comb}.

\section{Complexity and Performance Analysis \label{sec:Complexity-and-Performance}}

\subsection{Complexity Evaluation}

In this subsection, we briefly analyze both the computational cost
and memory requirements of the proposed packet combining schemes.
First, note that both algorithms have identical implementations. The
only difference comes from steps Table. \ref{tab:ITEQ_algorithm}.
\textbf{1.1.}, and Table. \ref{tab:Symbol_level_comb}. \textbf{1.1.3.}
Therefore, both techniques approximately have the same implementation
cost. In the following, we focus on the number of arithmetic additions
and memory required to perform recursions (\ref{eq:y_update}), (\ref{eq:D_update}),
and (\ref{eq:LLR_Comb}).

The main idea in the proposed algorithms is to exploit the diversity
available in multiple transmissions without explicitly storing required
soft channel outputs (i.e., signals and CFRs) or decisions (i.e.,
filter outputs), corresponding to all ARQ rounds. This is performed
with the aid of recursions (\ref{eq:y_update}), (\ref{eq:D_update}),
and (\ref{eq:LLR_Comb}), and translates into a memory requirement
of{\small{} }$2T_{c}N_{T}\left(N_{T}+1\right)$ and $T_{s}N_{T}2^{M}$
real values for chip-level and symbol-level turbo combining, respectively.
Note that in both schemes, the required memory size is insensitive
to the ARQ delay. The number of rounds only influences the number
of arithmetic additions required in the update procedures corresponding
to recursions (\ref{eq:y_update}), (\ref{eq:D_update}), and (\ref{eq:LLR_Comb}).
At each ARQ round, the chip-level turbo combining algorithm involves
$2T_{c}N_{T}\left(N_{T}+1\right)$ arithmetic additions to update
$\underline{\mathbf{\tilde{y}}}_{f}^{(k)}$ and $\underline{\mathbf{D}}_{i}^{(k)}$.
The symbol-level turbo combining scheme requires $T_{s}N_{T}N_{\mathrm{iter}}2^{M}$
arithmetic additions to update $\boldsymbol{\overline{\xi}}_{t,j}^{(k)}(s)$
at each round, where $N_{\mathrm{iter}}$ denotes the number of turbo
iterations. 

Table \ref{tab:SUMMARY-OF-ARITHMETIC} summarizes the maximum number
of arithmetic additions and memory size required by both schemes.
Note that the number of additions does not have a great impact on
receiver computational complexity. The required memory size is the
major implementation constraint to take into account when choosing
between chip-level and symbol-level combining. In the case of low-order
modulations (i.e., $M\leq2$), symbol-level has less memory requirements
than chip-level combining independently of the spreading factor $N$,
number of codes $C$, and number of transmit antennas $N_{T}$. For
high-order modulations, (i.e., $M\geq3$), the required memory size
mainly depends on system parameters. For instance, when $M=4$, $N_{T}=4$,
and the system is fully loaded, (i.e., $N=C$), chip-level combining
offers less memory requirements than symbol-level combining. When
the load factor is reduced to $50\%$, (i.e., $\frac{C}{N}=\frac{1}{2}$),
symbol-level becomes more attractive than chip-level.

\subsection{Performance Evaluation\label{sec:Performance-evaluation}}

In this subsection, we evaluate the throughput performance of the
proposed CP-CDMA MIMO ARQ turbo combining schemes. Following \cite{Caire_IT01},
we define the throughput as $\eta\triangleq\frac{\mathbb{{E}}\left[\mathcal{R}\right]}{\mathbb{{E}}\left[\mathcal{K}\right]}$,
where $\mathcal{R}$ is a random variable (RV) that takes $R$ when
the packet is correctly received or zero when the packet is erroneous
after $K$ ARQ rounds. $\mathcal{K}$ is a RV that denotes the number
of rounds used for transmitting one data packet. We use Monte Carlo
simulations for evaluating $\eta$. 

We consider a STC using a $\frac{1}{2}$-rate convolutional encoder
with polynomial generators $\left(35,23\right)_{8}$, quadrature phase
shift keying (QPSK) modulation, $N_{T}=2$ transmit antennas, and
a spreading factor $N=16$. The length of the code bit frame is $1024$
bits including tails. We evaluate the throughput performance for the
following loads: $25\%$ (i.e., $C=4$), $50\%$ (i.e., $C=8$), and
$100\%$ (i.e., $C=16$), which correspond to rates $R=8$, $R=16$,
and $R=32$, respectively. The ARQ delay is $K=3$. The broadband
MIMO channel has $L=10$ chip-spaced equal power taps, and the CP
length is $T_{CP}=10$. The $E_{c}/N_{0}$ ratio appearing in all
figures is the signal to noise ratio (SNR) per chip per receive antenna.
We use Max-Log-maximum \emph{a posteriori} (MAP) for SISO decoding.
The number of turbo iterations is set to three. In all scenarios,
we consider the matched filter bound (MFB) throughput performance
of the corresponding CP-CDMA MIMO ARQ channel to evaluate the ICI
cancellation capability achieved by the proposed techniques. 

In Fig. \ref{fig:BLER_2_2}, we report throughput performance curves
for a balanced MIMO configuration, i.e., $N_{R}=N_{T}=2$. We observe
that both combining schemes have similar throughput performance for
quarter and half loads. In the case of full load, chip-level combining
outperforms symbol-level combining in the region of low SNR. For instance,
the performance gap is around $0.6$dB at $\eta=12.5$bit/s/Hz throughput.
Also, note that for all configurations, the slopes of the throughput
curves of both techniques are asymptotically similar to that of the
MFB. Therefore, both combining schemes asymptotically achieve the
diversity order of the corresponding CP-CDMA MIMO ARQ channel. 

In Fig. \ref{fig:BLER2_1}, we provide throughput curves when only
one receive antenna ($N_{R}=1$) is used, i.e., unbalanced MIMO configuration.
In this scenario, chip-level combining clearly outperforms symbol-level
combining for half and full loads. The performance gap is about $3$dB
at $\eta=12.5$bit/s/Hz for a full load configuration. This suggests
that chip-level turbo combining can be used for high speed downlink
CDMA MIMO systems with high loads. Note that, both techniques fail
to achieve the full diversity order in the case of half and full loads.

\section{Conclusions\label{sec:Conclusions}}

In this paper, efficient turbo receiver schemes for multi-code CP-CDMA
transmission with ARQ operating over broadband MIMO channel were investigated.
Two packet combining algorithms were introduced. The chip-level technique
performs packet combining jointly with chip-level MMSE FDE. The symbol-level
scheme combines multiple transmissions at the level of the soft demapper.
We analyzed the complexity and memory size required by both techniques,
and showed that, from an implementation point of view, chip-level
is more attractive than symbol-level combining for systems with high
modulation order and load factor (number of codes with respect to
the spreading factor). We also investigated the throughput performance.
Simulations demonstrated that both techniques approximately have similar
performance for balanced MIMO configurations. In the case of unbalanced
configurations (more transmit than receive antennas), chip-level combining
outperforms symbol-level combining especially for full load factors.

\begin{center}
\begin{figure}[H]
\noindent \begin{centering}
\includegraphics[scale=0.8]{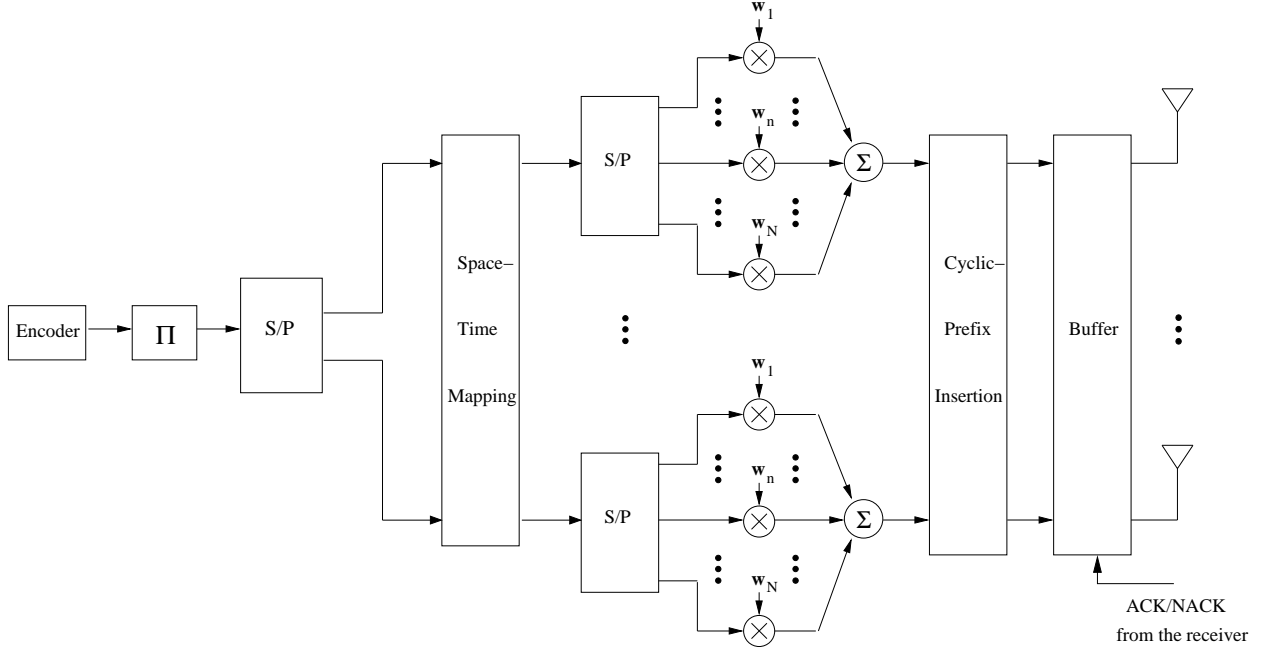} 
\par\end{centering}

\caption{{\footnotesize \label{fig:STBICM_ARQ}CP-CDMA MIMO transmission scheme
with ACK/NACK.}}

\end{figure}
\vspace{3cm}

\par\end{center}

\begin{center}
\begin{table}[H]
\caption{{\footnotesize \label{tab:ITEQ_algorithm}}\textsc{\footnotesize Summary
Of The Chip-Level Turbo Combining Algorithm}{\footnotesize{} }}

\begin{centering}
\begin{tabular}{llll}
\hline 
\multicolumn{4}{l}{}\tabularnewline
\textbf{\small 0.} & \multicolumn{3}{l}{\textbf{\small Initialization}}\tabularnewline
 & \multicolumn{3}{l}{{\small Initialize $\underline{\mathbf{\tilde{y}}}_{f}^{(0)}$ and
$\underline{\mathbf{D}}_{i}^{(0)}$ with $\boldsymbol{0}_{T_{c}N_{T}\times1}$
and $\boldsymbol{{0}}_{N_{T}\times N_{T}}$, respectively.}}\tabularnewline
\textbf{\small 1.} & \multicolumn{3}{l}{\textbf{\small Combining at round $k$}}\tabularnewline
 & \textbf{\small 1.1.} & \multicolumn{2}{l}{{\small Update $\underline{\mathbf{\tilde{y}}}_{f}^{(k)}$ and }$\underline{\mathbf{D}}_{i}^{(k)}${\small{}
according to (\ref{eq:y_update}) and (\ref{eq:D_update}).}}\tabularnewline
 & \textbf{\small 1.2.} & \multicolumn{2}{l}{{\small At each iteration,}}\tabularnewline
 &  & \textbf{\small 1.2.1} & {\small Compute the forward and backward filters using (\ref{eq:forward_filter})
and (\ref{eq:backward_filter}).}\tabularnewline
 &  & \textbf{\small 1.2.2} & {\small Compute the MMSE estimate of $\mathbf{x}_{f}$ using (\ref{eq:MMSE_plus_IC_metric_FD}).}\tabularnewline
 &  & \textbf{\small 1.2.3} & {\small Compute extrinsic LLRs $\phi_{t,j,m}^{(e)}$ according to
(\ref{eq:Ext_LLR}).}\tabularnewline
 & \textbf{\small 1.3.} & \multicolumn{2}{l}{{\small end }\textbf{\small 1.2.}}\tabularnewline
\multicolumn{4}{l}{}\tabularnewline
\hline
\end{tabular}
\par\end{centering}
\end{table}

\par\end{center}

\begin{center}
\newpage{}%
\begin{table}[H]
\caption{{\small \label{tab:Symbol_level_comb}}\textsc{\footnotesize Summary
Of The Symbol-Level Turbo Combining Algorithm}}

\begin{centering}
\begin{tabular}{llll}
\hline 
\multicolumn{4}{l}{}\tabularnewline
\textbf{\small 0.} & \multicolumn{3}{l}{\textbf{\small Initialization:}}\tabularnewline
 & \multicolumn{3}{l}{{\small Initialize $\boldsymbol{\overline{\xi}}_{t,j}^{(0)}(s)$ with
$0$.}}\tabularnewline
\textbf{\small 1.} & \multicolumn{3}{l}{\textbf{\small Combining at round $k$}}\tabularnewline
 & \textbf{\small 1.1.} & \multicolumn{2}{l}{{\small At each iteration,}}\tabularnewline
 &  & \textbf{\small 1.1.1} & {\small Compute the forward and backward filters using (\ref{eq:forward_filter})
and (\ref{eq:backward_filter}) with $\underline{\mathbf{D}}_{i}^{(k)}=\boldsymbol{{\Lambda}}_{i}^{\left(k\right)^{H}}\boldsymbol{{\Lambda}}_{i}^{\left(k\right)}$.}\tabularnewline
 &  & \textbf{\small 1.1.2} & {\small Compute the MMSE estimate on $\mathbf{x}_{f}$ using (\ref{eq:MMSE_plus_IC_metric_FD})
and $\underline{\mathbf{\tilde{y}}}_{f}^{(k)}=\boldsymbol{{\Lambda}}^{\left(k\right)^{H}}\mathbf{y}_{f}^{\left(k\right)}$.}\tabularnewline
 &  & \textbf{\small 1.1.3} & {\small Update $\boldsymbol{\overline{\xi}}_{t,j}^{(k)}(s)$ according
to (\ref{eq:LLR_Comb}).}\tabularnewline
 &  & \textbf{\small 1.1.4} & {\small Compute extrinsic LLRs $\phi_{t,j,m}^{(e)}$ using (\ref{eq:Symbol_Comb}).}\tabularnewline
 & \textbf{\small 1.3.} & \multicolumn{2}{l}{{\small end }\textbf{\small 1.1.}}\tabularnewline
\multicolumn{4}{l}{}\tabularnewline
\hline
\end{tabular}
\par\end{centering}
\end{table}

\par\end{center}

\begin{center}
\vspace{7cm}
\begin{table}[H]
\caption{{\small \label{tab:SUMMARY-OF-ARITHMETIC}}\textsc{\footnotesize Summary
of the Maximum Number of Arithmetic Additions, and Memory Size}}

\begin{centering}
\begin{tabular}{ccc}
 & \textbf{\small Chip-Level Combining } & \textbf{\small Symbol-Level Combining }\tabularnewline
\hline
\hline 
\textbf{\small Arithmetic Additions} & {\scriptsize $2T_{c}N_{T}\left(K-1\right)\left(N_{T}+1\right)$ } & {\scriptsize $T_{s}N_{T}\left(K-1\right)N_{\mathrm{iter}}2^{M}$}\tabularnewline
\textbf{\small Memory} & {\scriptsize $2T_{c}N_{T}\left(N_{T}+1\right)$ } & {\scriptsize $T_{s}N_{T}2^{M}$ }\tabularnewline
\hline
\end{tabular}
\par\end{centering}
\end{table}

\par\end{center}

\begin{center}
\newpage{}%
\begin{figure}[H]
\begin{centering}
\includegraphics[scale=0.8]{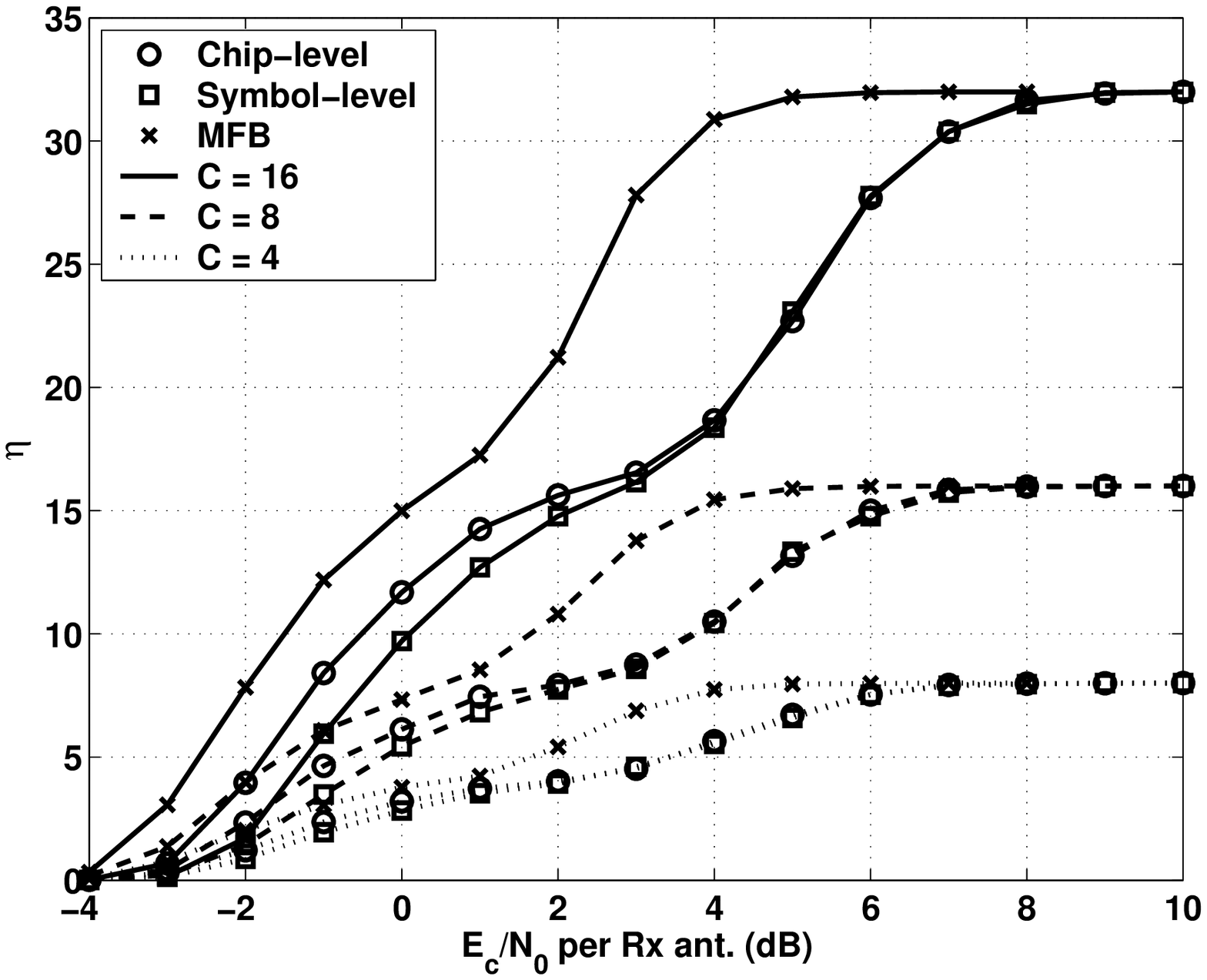}
\par\end{centering}

\begin{centering}

\par\end{centering}

\caption{{\small \label{fig:BLER_2_2}}{\footnotesize Throughput performance
with $N_{T}=2$, $N_{R}=2$, $L=10$ equal power tap profile.}}

\end{figure}

\par\end{center}

\begin{center}
\newpage{}%
\begin{figure}[H]
\begin{centering}
\includegraphics[scale=0.8]{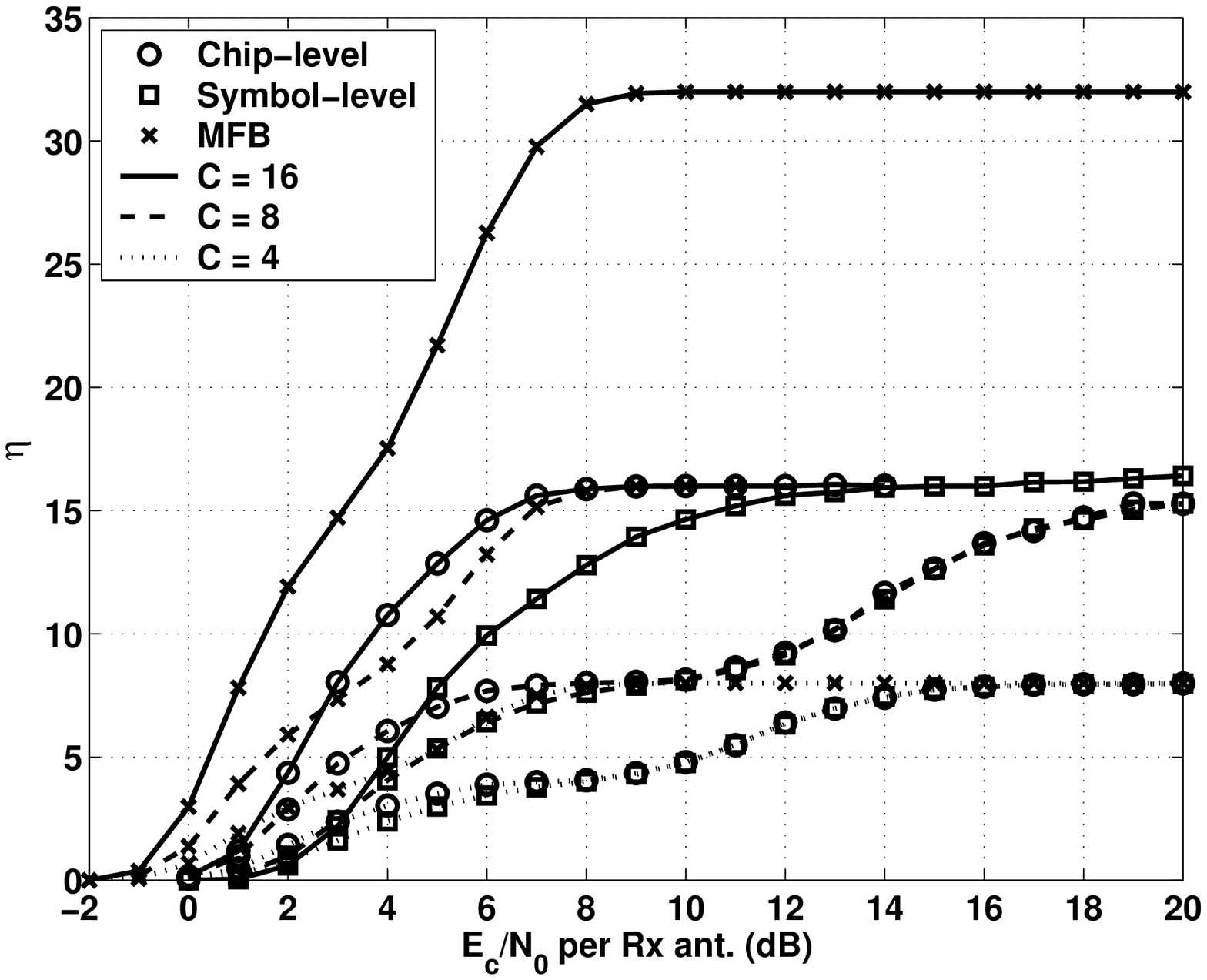}
\par\end{centering}

\begin{centering}

\par\end{centering}

\caption{{\small \label{fig:BLER2_1}}{\footnotesize Throughput performance
with $N_{T}=2$, $N_{R}=1$, $L=10$ equal power tap profile.}}

\end{figure}

\par\end{center}
\end{document}